\documentclass{article}
\usepackage{fullpage}
\usepackage[american]{babel}
\usepackage{amssymb}
\usepackage{algorithmic}
\usepackage{algorithm}
\usepackage{multirow}
\usepackage{graphicx}

\newcommand{\eqn}[1]{(\ref{#1})}
\newcommand{\beql}[1]{\begin{equation}\label{#1}}
\newcommand{\eeq}{\end{equation}}
\newtheorem{theo}{Theorem}
\newtheorem{defi}{Definiton}
\newtheorem{asm}{Assumption}

\title{An efficient dynamic programming algorithm for the generalized LCS problem with multiple substring inclusive constraints}
\author{Daxin Zhu, Lei Wang, Yingjie Wu, and Xiaodong Wang}
\begin{document}
\maketitle

\begin{abstract}
In this paper, we consider a generalized longest common subsequence problem with multiple substring inclusive constraints. For the two input sequences $X$ and $Y$ of lengths $n$ and $m$, and a set of $d$ constraints $P=\{P_1,\cdots,P_d\}$ of total length $r$, the problem is to find a common subsequence $Z$ of $X$ and $Y$ including each of constraint string in $P$ as a substring and the length of $Z$ is maximized.
A new dynamic programming solution to this problem is presented in this paper. The correctness of the new algorithm is proved. The time complexity of our algorithm is $O(d2^dnmr)$. In the case of the number of constraint strings is fixed, our new algorithm for the generalized longest common subsequence problem with multiple substring inclusive constraints requires $O(nmr)$ time and space.
\end{abstract}

\section{Introduction}
The longest common subsequence (LCS) problem is a classic computer science problem, and has applications in bioinformatics. It is further widely applied in diverse areas, such as file comparison, pattern matching and computational biology\cite{3,4,8,9}. Given two sequences $X$ and $Y$, the longest common subsequence problem is to find a subsequence of $X$ and $Y$ whose length is the longest among all common subsequences of the two given sequences. It differs from the problems of finding common substrings: unlike substrings, subsequences are not required to occupy consecutive positions within the original sequences.
The most referred algorithm, proposed by Wagner and Fischer \cite{29}, solves the LCS problem by using a dynamic programming algorithm in quadratic time. Other advanced algorithms were proposed in the past decades \cite{2,3,4,16,17,19,21}.
If the number of input sequences is not fixed, the problem to find the LCS of multiple sequences has been proved to be NP-hard \cite{23}. Some approximate and heuristic algorithms were proposed for these problems \cite{6,25}.

For some biological applications some constraints must be applied to the LCS problem. These kinds of variants of the LCS problem are called the constrained LCS (CLCS) problem. One of the recent variants of the LCS problem, the constrained longest common subsequence (CLCS) which was first addressed by Tsai \cite{27}, has received much attention.
It generalizes the LCS measure by introducing of a third sequence, which allows to extort that the obtained CLCS has some special properties \cite{26}.
For two given input sequences $X$ and $Y$ of lengths $m$ and $n$, respectively, and a constrained sequence $P$ of length $r$, the CLCS problem is to find the common subsequences $Z$ of $X$ and $Y$ such that $P$ is a subsequence of $Z$ and the length of $Z$ is the maximum.
The most referred algorithms were proposed independently \cite{5,8}, which solve the CLCS problem in $O(mnr)$ time and space by using dynamic programming algorithms. Some improved algorithms have also been proposed \cite{11,18}. The LCS and CLCS problems on the indeterminate strings were discussed in \cite{20}.
Moreover, the problem was extended to the one with weighted constraints, a more generalized problem \cite{24}.

Recently, a new variant of the CLCS problem, the restricted LCS problem, was proposed \cite{14}, which excludes the given constraint as a subsequence of the answer. The restricted LCS problem becomes NP-hard when the number of constraints is not fixed.
Some more generalized forms of the CLCS problem, the generalized constrained longest common subsequence (GC-LCS) problems, were addressed independently by Chen and Chao \cite{7}.
For the two input sequences $X$ and $Y$ of lengths $n$ and $m$, respectively, and a constraint string $P$ of length $r$, the GC-LCS problem is a set of four problems which are to find the LCS of $X$ and $Y$ including/excluding $P$ as a subsequence/substring, respectively. The four generalized constrained LCS\cite{7} can be summarized in Table 1.

\begin{table}[ht]
\begin{center}
\caption{The GC-LCS problems}
\begin{tabular}{|l|l|l|}
\hline\hline
Problem & Input & Output\\\hline
\multirow{2}{*}{SEQ-IC-LCS} & \multirow{2}{*}{$X$,$Y$, and $P$} & The longest common subsequence of $X$ and $Y$ \\
 &  & including $P$ as a subsequence \\\hline
\multirow{2}{*}{STR-IC-LCS} & \multirow{2}{*}{$X$,$Y$, and $P$} & The longest common subsequence of $X$ and $Y$ \\
 &  & including $P$ as a substring \\\hline
\multirow{2}{*}{SEQ-EC-LCS} & \multirow{2}{*}{$X$,$Y$, and $P$} & The longest common subsequence of $X$ and $Y$ \\
 &  & excluding $P$ as a subsequence\\\hline
\multirow{2}{*}{STR-EC-LCS} & \multirow{2}{*}{$X$,$Y$, and $P$} & The longest common subsequence of $X$ and $Y$ \\
 &  & excluding $P$ as a substring\\\hline
\end{tabular}
\end{center}
\end{table}

For the four problems in Table 1, $O(mnr)$  time algorithms were proposed \cite{7}.
For all four variants in Table 1, $O(r(m + n) + (m + n) \log(m+n))$ time algorithms were proposed by using the finite automata \cite{12}.
Recently, a quadratic algorithm to the STR-IC-LCS problem was proposed \cite{10}, and the time complexity of \cite{12} was pointed out not correct.

The four GC-LCS problems can be generalized further to the cases of multiple constraints. In these generalized cases, the single constrained pattern $P$ will be generalized to a set of $d$ constraints $P=\{P_1,\cdots,P_d\}$ of total length $r$, as shown in Table 2.

\begin{table}[ht]
\begin{center}
\caption{The Multiple-GC-LCS problems}
\begin{tabular}{|l|l|l|}
\hline\hline
Problem & Input & Output\\\hline
\multirow{2}{*}{M-SEQ-IC-LCS} & $X$,$Y$, and a set of constraints & The longest common subsequence of $X$ and $Y$ \\
 &  $P=\{P_1,\cdots,P_d\}$  & including each of constraint $P_i\in P$ as a subsequence \\\hline
\multirow{2}{*}{M-STR-IC-LCS} & $X$,$Y$, and a set of constraints & The longest common subsequence of $X$ and $Y$ \\
 &  $P=\{P_1,\cdots,P_d\}$  & including each of constraint $P_i\in P$ as a substring \\\hline
\multirow{2}{*}{M-SEQ-EC-LCS} & $X$,$Y$, and a set of constraints & The longest common subsequence of $X$ and $Y$ \\
 &  $P=\{P_1,\cdots,P_d\}$  & excluding each of constraint $P_i\in P$ as a subsequence\\\hline
\multirow{2}{*}{M-STR-EC-LCS} & $X$,$Y$, and a set of constraints & The longest common subsequence of $X$ and $Y$ \\
 &  $P=\{P_1,\cdots,P_d\}$  & excluding each of constraint $P_i\in P$ as a substring\\\hline
\end{tabular}
\end{center}
\end{table}

The problem M-SEQ-IC-LCS has been proved to be NP-hard in \cite{13}.
The problem M-SEQ-EC-LCS has also been proved to be NP-hard in \cite{14,28}.
In addition, the problems M-STR-IC-LCS and M-STR-EC-LCS were also declared to be NP-hard in \cite{7}, but without a proof.
The exponential-time algorithms for solving these two problems were also presented in \cite{7}.

We will discuss the problem M-STR-IC-LCS in this paper.
The failure functions in the Knuth-Morris-Pratt algorithm \cite{22} for solving the string matching problem have been proved very helpful for solving the STR-IC-LCS problem. It has been found by Aho and Corasick\cite{1} that the failure functions can be generalized to the case of keyword tree to speedup the exact string matching of multiple patterns. This idea can be very helpful in our dynamic programming algorithm. This is the principle  idea of our new algorithm.

The organization of the paper is as follows.

In the following 4 sections, we describe our presented dynamic programming algorithm for the M-STR-IC-LCS problem.

In Section 2 the preliminary knowledge for presenting our algorithm for the M-STR-IC-LCS problem is discussed.
In Section 3 we give a new dynamic programming solution for the M-STR-IC-LCS problem with time complexity $O(d2^dnmr)$, where $n$ and $m$ are the lengths of the two given input strings, and $r$ is the total length of $d$ constraint strings.
In Section 4, we discuss the issues to implement the algorithm efficiently.
Some concluding remarks are provided in Section 5.

\section{Preliminaries}
A sequence is a string of characters over an alphabet $\sum$. A subsequence of a sequence $X$ is obtained by deleting zero or more characters from $X$ (not necessarily contiguous). A substring of a sequence $X$ is a subsequence of successive characters within $X$.

For a given sequence $X=x_1x_2\cdots x_n$ of length $n$, the $i$th character of $X$ is denoted as $x_i \in \sum$ for any $i=1,\cdots,n$. A substring of $X$ from position $i$ to $j$ can be denoted as $X[i:j]=x_ix_{i+1}\cdots x_j$. If $i\neq 1$ or $j\neq n$, then the substring $X[i:j]=x_ix_{i+1}\cdots x_j$ is called a proper substring of $X$. A substring $X[i:j]=x_ix_{i+1}\cdots x_j$ is called a prefix or a suffix of $X$ if $i=1$ or $j=n$, respectively.

For the two input sequences $X=x_1x_2\cdots x_n$ and $Y=y_1y_2\cdots y_m$ of lengths $n$ and $m$, respectively, and a set of $d$ constraints $P=\{P_1,\cdots,P_d\}$ of total length $r$, the problem M-STR-IC-LCS is to find an LCS of $X$ and $Y$ including each of constraint $P_i\in P$ as a substring.

Keyword tree (Aho-Corasick Automaton)\cite{1,9,15} is a main data structure in our dynamic programming algorithm to process the constraint set $P$ of the M-STR-IC-LCS problem.

\begin{defi}\label{df1}
The keyword tree for set $P$ is a rooted directed tree $T$ satisfying 3 conditions: 1. each edge is labeled with exactly one character; 2. any two edges out of the same node have distinct labels; and 3. every string $P_i$ in $P$ maps to some node $v$ of $T$ such that the characters on the path from the root of $T$ to $v$ exactly spell out $P_i$, and every leaf of $T$ is mapped to some string in $P$.
\end{defi}

\begin{figure}
\centering
\includegraphics[width=5cm,height=4cm]{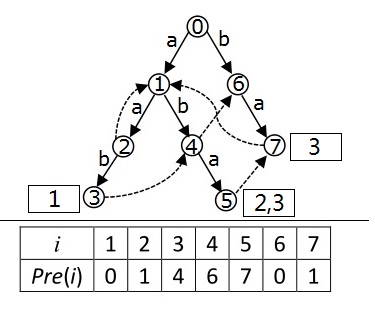}
\caption{Keyword Trees}
\end{figure}

In order to identify the nodes of $T$, we assign numbers $0,1,\cdots,t-1$ to all $t$ nodes of $T$ in their preorder numbering. Then, each node will be assigned an integer $i,0\leq i<t$, as shown in Fig.1.
For each node numbered $i$ of a keyword tree $T$, the concatenation of characters on the path from the root to the node $i$ spells out a string denoted as $L(i)$. The string $L(i)$ is also called the label of the node $i$ in the keyword tree $T$.
For example, Fig.1 shows the keyword tree $T$ for the constraint set $P=\{aab,aba,ba\}$, where $P_1=aab,P_2=aba,P_3=ba$, and $d=3,r=8$.
Clearly, every node in the keyword tree corresponds to a prefix of one of the strings in set $P$, and every prefix of a string $P_i$ in $P$ maps to a distinct node in the keyword tree $T$.
The keyword tree for set $P$ of total length $r$ of all strings can be easily constructed in $O(r)$ time for a constant alphabet size.

The keyword tree can be extended into an automaton, Aho-Corasick automaton, which consists of three functions, a goto function, an output function and a failure function. The goto function is represented as the solid edges of the keyword tree and the output function indicates when the matches occur and which strings are output. For each node $i$, its output function is denoted as $O_i$, a set of indices which indicates when the node $i$ is reached then for each index $j\in O_i$, the string $P_j$ is matched. For example, the output sets of nodes 3,5 and 7 are $O_3=\{1\}$,$O_5=\{2,3\}$ and $O_7=\{3\}$, which means that the outputs of node 3,5 and 7 are $\{P_1=aab\}$,$\{P_2=aba,P_3=ba\}$ and $\{P_3=ba\}$, respectively.

The failure function indicates which node to go if there is no character to be further matched. It is a generalization of the failure functions in the Knuth-Morris-Pratt algorithm for solving the string matching problem. It is represented by the dashed edges in Fig.1.

For any node $i$ of $T$, define $lp(i)$ to be the length of the longest proper suffix of string $L(i)$ that is a prefix of some string in $T$.
It can be verified readily that for each node $i$ of $T$, if $A$ is an $lp(i)$-length suffix of string $L(i)$, then there must be a unique node $pre(i)$ in $T$ such that $L(pre(i))=A$. If $lp(i)=0$ then $pre(i)=0$ is the root of $T$.

%\begin{defi}\label{df2}
The ordered pair $(i,pre(i))$ is called a failure link.
%\end{defi}
The failure link is a direct generalization of the failure functions in the KMP algorithm.
For example, in Fig.1, failure links are shown as pointers from every node $i$ to node $pre(i)$ where $lp(i)>0$. The other failure links point to the root and are not shown.
The failure links of $T$ define actually a failure function $pre$ for the constraint set $P$.
As stated in \cite{1,9}, for a constant alphabet size, in the worst case, the failure function $pre$ can be computed in $O(r)$ time.

The failure list of a given node is the ordered list of the nodes which locate on the path to the root via dashed edges.
For example, for the nodes $i=1,2,3,4,5,6,7$ , the corresponding values of failure function are $pre(i)=0,1,4,6,7,0,1$.
The failure list of node 5 is $\{7\rightarrow 1\rightarrow 0\}$, and the failure list of node 6 is $\{0\}$, as shown in Fig.1.

The failure function $pre$ is used to speedup the search for all occurrences in a text $Z$ of strings from $P$.
For each node $i$ of $T$, and a character $c\in \sum$, if no edges out of the node $i$ is labeled $c$, then the failure link of node $i$ direct the search to the node $pre(i)$. It is equivalent to add the edge $(i,pre(i))$ labeled $c$ to the node $i$. This set matching method generalized the next function in KMP algorithm to the Aho-Corasick-next function as follows.

\begin{defi}\label{df2}
Given a keyword tree $T$ and its failure function, for each node $i$ of $T$ and each character $c\in \sum$, Aho-Corasick-next function $\delta(i,c)$ denotes the destination of the first node in $i$'s failure list which has an edge labeled $c$. If there exists no such node in the failure list, the function returns the root.
\end{defi}

Table 3 shows the Aho-Corasick-next function $\delta$ corresponding to the example in Fig.1.

\begin{table}[ht]
\begin{center}
\caption{Aho-Corasick-next function}
\begin{tabular}{|c|c|c|c|c|c|c|c|c|}
\hline\hline
$\delta$&0&1&2&3&4&5&6&7\\\hline
a&1&2&1&4&5&1&7&1\\\hline
b&6&4&3&0&0&1&0&1\\\hline
\end{tabular}
\end{center}
\end{table}

We take node 4 as an example. It can be seen from Fig.1 that $\delta(4,a)=5$ and $\delta(4,b)=0$.
It is easy to see that each element of Aho-Corasick-next function can be computed in constant time.

The symbol $\oplus$ is also used to denote the string concatenation. For example, if $S_1=aaa$ and $S_2=bbb$, then it is readily seen that  $S_1\oplus S_2=aaabbb$.

\section{Our Main Result: A Dynamic Programming Algorithm}
Let $T$  be a keyword tree for the given constraint set $P$, and $Z[1:l]=z_1,z_2,\cdots,z_l$ be any common subsequence of $X$ and $Y$. If we search the set matching of $Z$ from the root of $T$ in the direction of the
Aho-Corasick-next function $\delta$ of $T$, then the search will stop in a node $i$ of $T$. All such common subsequence of $X$ and $Y$ can be classified into a group $i$, $0\leq i<t$. These $t$ groups are still not sufficient to distinguish the different states in our dynamic programming algorithm, since the common subsequence of $X$ and $Y$ in the same group may contain different subset of $P$. Therefore, we must divide each group into $2^d$ new states by attaching $d$ flags to denote the combinations which constraints have been kept. The $d$ flags can be record by a $d$ bits vector $s$. If the string $P_j\in P$ is kept, then the bit $j$ of $s$ is set to 1, otherwise 0. There are total $2^d$ different such bit vectors, denoted as $s_0,s_1,\cdots, s_{2^d-1}$ as follows.

\begin{defi}\label{df3}
$\newline$
\begin{itemize}
\item Let $0\leq j<2^d$, and $j=\sum_{i=1}^d b_i2^{i-1}$. Then the set $s_j$ is defined as $s_j=\{i \mid b_i=1,1\leq i\leq d\}$.

\item If a subset of strings $s=\{P_{k_1},P_{k_2},\cdots,P_{k_h}\}\subseteq P$ must be added to the set $s_j$, then the set $s_j$ becomes $s_k$, where $k=j\bigvee \sum_{i=1}^h 2^{k_i-1}$, and the operation $\bigvee$ is a bitwise or operation of two integers. In this case we denote $s_k=s_j\bigcup s$.

\item For a sequence $z$ with state $(\alpha,\beta)$ in a given keyword tree $T$, and a character $c\in \sum$, we now consider the state of the sequence $\bar{z}=z\bigoplus c$ in $T$.
From the node $\alpha$, the search for $\bar{z}$ will go to node $\bar{\alpha}=\delta(\alpha,c)$. If $O_{\bar{\alpha}}$, the output set of the node $\bar{\alpha}$ is not empty, then the strings of $O_{\bar{\alpha}}$ must be included in the sequence $\bar{z}$, and thus the set $s_\beta$ will changed to $s_{\bar{\beta}}=s_\beta\bigcup O_{\bar{\alpha}}$. In this case we denote $\bar{\beta}=\gamma(\alpha,\beta,c)$.
In other words, the state of the $\bar{z}=z\bigoplus c$ in $T$ becomes $(\delta(\alpha,c),\gamma(\alpha,\beta,c))$.
\end{itemize}
\end{defi}

For example, in the example of Fig.1, we have $d=3$, and $s_1=\{1\}$, $s_6=\{2,3\}$, $s_7=\{1,2,3\}$, $s_7=s_1\bigcup s_6$.

Finally we have $t2^d$ different states in our dynamic programming algorithm. For each pair $(i,j)$, $0\leq i<t, 0\leq j<2^d$, the state $(i,j)$ represents the set of common subsequence of $X$ and $Y$ in group $i$ and the subset of $P$ contained in the subsequence is recorded by bit vector $s_j$.

\begin{defi}\label{df4}
Let $Z(i,j,(\alpha,\beta))$ denote the set of all LCSs of $X[1:i]$ and $Y[1:j]$ with state $(\alpha,\beta)$, where $1\leq i\leq n, 1\leq j\leq m$, and $0\leq \alpha<t, 0\leq \beta<2^d$.
The length of an LCS in $Z(i,j,(\alpha,\beta))$ is denoted as $f(i,j,(\alpha,\beta))$.
\end{defi}

If we can compute $f(i,j,(\alpha,\beta))$ for any $1\leq i\leq n, 1\leq j\leq m$, and $0\leq \alpha<t, 0\leq \beta<2^d$ efficiently, then the length of an LCS of $X$ and $Y$ including $P$ must be $\max\limits_{0\leq i<t}\left\{f(n,m,(i,2^d-1))\right\}$.

By using the keyword tree data structure described in the last section, we can give a recursive formula for computing $f(i,j,(\alpha,\beta))$ by the following Theorem.

\begin{theo}\label{th1}
For the two input sequences $X=x_1x_2\cdots x_n$ and $Y=y_1y_2\cdots y_m$ of lengths $n$ and $m$, respectively, and a set of $d$ constraints $P=\{P_1,\cdots,P_d\}$ of total length $r$, let $Z(i,j,(\alpha,\beta))$ and $f(i,j,(\alpha,\beta))$ be defined as in Definition \ref{df4}.
Suppose a keyword tree $T$ for the constraint set $P$ has been built, and the $t$ nodes of $T$ are numbered in their preorder numbering.
The label of the node numbered $k(0\leq k<t)$ is denoted as $L(k)$.
Then, for any $1\leq i\leq n, 1\leq j\leq m$, and $0\leq \alpha<t, 0\leq \beta<2^d$, $f(i,j,(\alpha,\beta))$ can be computed by the following recursive formula \eqn{eq31}.

%\begin{figure}[bt]
\beql{eq31}
f(i,j,(\alpha,\beta))=\left\{\begin{array}{ll}
\max\left\{ f(i-1,j,(\alpha,\beta)),f(i,j-1,(\alpha,\beta)) \right\} & \texttt{if } x_i\neq y_j,\\
\max\left\{
f(i-1,j-1,(\alpha,\beta)),1+\max\limits_{(\bar{\alpha},\bar{\beta})\in S(\alpha,\beta,x_i)}\left\{f(i-1,j-1,(\bar{\alpha},\bar{\beta}))\right\}
\right\} & \texttt{if } x_i= y_j.
\end{array} \right.
\eeq
%\end{figure}

Where,
\beql{eq32}
S(\alpha,\beta,x_i)=\{(\bar{\alpha},\bar{\beta})|0\leq \bar{\alpha}<t, 0\leq \bar{\beta}<2^d,\delta(\bar{\alpha},x_i)=\alpha,\gamma(\bar{\alpha},\bar{\beta},x_i)=\beta\}
\eeq

The boundary conditions of this recursive formula are $f(i,0,(0,0)) = f(0,j,(0,0)) = 0$ for any $0\leq i\leq n, 0\leq j\leq m$.
\end{theo}

\noindent{\bf Proof.}

For any $0\leq i\leq n, 0\leq j\leq m$, and $0\leq \alpha<t, 0\leq \beta<2^d$, suppose $f(i,j,(\alpha,\beta))=l$ and $z=z_1 \cdots z_l\in Z(i,j,(\alpha,\beta))$.

First of all, we notice that for each pair $(i',j'), 1\leq i'\leq n, 1\leq j'\leq m$, such that $i'\leq i$ and $j'\leq j$, we have $f(i',j',(\alpha,\beta)) \leq f(i,j,(\alpha,\beta))$, since a common subsequence $z$ of $X[1:i']$ and $Y[1:j']$ with state $(\alpha,\beta)$ is also a common subsequence of $X[1:i]$ and $Y[1:j]$ with state $(\alpha,\beta)$.

(1) In the case of $x_i\neq y_j$, we have $x_i\neq z_l$ or $y_j\neq z_l$.

(1.1)If $x_i\neq z_l$, then $z=z_1 \cdots z_l$ is a common subsequence of $X[1:i-1]$ and $Y[1:j]$ with state $(\alpha,\beta)$, and so $f(i-1,j,(\alpha,\beta)) \geq l$. On the other hand, $f(i-1,j,(\alpha,\beta))\leq f(i,j,(\alpha,\beta)) = l$. Therefore, in this case we have $f(i,j,(\alpha,\beta)) = f(i-1,j,(\alpha,\beta))$.

(1.2)If $y_j\neq z_l$, then we can prove similarly that in this case, $f(i,j,(\alpha,\beta)) = f(i,j-1,(\alpha,\beta))$.

Combining the two subcases we conclude that in the case of $x_i\neq y_j$, we have $$f(i,j,(\alpha,\beta))=\max\left\{ f(i-1,j,(\alpha,\beta)),f(i,j-1,(\alpha,\beta)) \right\}.$$

(2) In the case of $x_i=y_j$, there are also two cases to be distinguished.

(2.1)If $x_i=y_j\neq z_l$,  then $z=z_1 \cdots z_l$ is also a common subsequence of $X[1:i-1]$ and $Y[1:j-1]$ with state $(\alpha,\beta)$, and so $f(i-1,j-1,(\alpha,\beta)) \geq l$. On the other hand, $f(i-1,j-1,(\alpha,\beta))\leq f(i,j,(\alpha,\beta)) = l$. Therefore, in this case we have $f(i,j,(\alpha,\beta)) = f(i-1,j-1,(\alpha,\beta))$.

(2.2)If $x_i=y_j=z_l$, then $f(i,j,(\alpha,\beta)) = l>0$ and $z=z_1 \cdots z_l$ is an LCS of $X[1:i]$ and $Y[1:j]$ with state $(\alpha,\beta)$.

Let the state of $(z_1,\cdots, z_{l-1})$ be $(\bar{\alpha},\bar{\beta})$, then we have $(\bar{\alpha},\bar{\beta})\in S(\alpha,\beta,x_i)$, since $z_l=x_i$.
It follows that $z_1 \cdots z_{l-1}$ is a common subsequence of $X[1:i-1]$ and $Y[1:j-1]$ with state $(\bar{\alpha},\bar{\beta})$.
Therefore, we have

%\beql{eq32}
\[
f(i-1,j-1,(\bar{\alpha},\bar{\beta}))\geq l-1
\]
%\eeq

Furthermore, we have

\[
\max\limits_{(\bar{\alpha},\bar{\beta})\in S(\alpha,\beta,x_i)}\left\{f(i-1,j-1,(\bar{\alpha},\bar{\beta}))\right\}\geq l-1
\]

In other words,
\beql{eq34}
f(i,j,(\alpha,\beta))\leq 1+\max\limits_{(\bar{\alpha},\bar{\beta})\in S(\alpha,\beta,x_i)}\left\{f(i-1,j-1,(\bar{\alpha},\bar{\beta}))\right\}
\eeq

On the other hand, for any $(\bar{\alpha},\bar{\beta})\in S(\alpha,\beta,x_i)$, and $v=v_1 \cdots v_h\in Z(i-1,j-1,(\bar{\alpha},\bar{\beta}))$, $v\oplus x_i$ is a common subsequence of $X[1:i]$ and $Y[1:j]$ with state $(\alpha,\beta)$.
Therefore, $f(i,j,(\alpha,\beta))=l\geq 1+h=1+f(i-1,j-1,(\bar{\alpha},\bar{\beta}))$, and so we conclude that,
\beql{eq35}
f(i,j,(\alpha,\beta))\geq 1+\max\limits_{(\bar{\alpha},\bar{\beta})\in S(\alpha,\beta,x_i)}\left\{f(i-1,j-1,(\bar{\alpha},\bar{\beta}))\right\}
\eeq

Combining \eqn{eq34} and \eqn{eq35} we have, in this case,

\beql{eq36}
f(i,j,(\alpha,\beta))= 1+\max\limits_{(\bar{\alpha},\bar{\beta})\in S(\alpha,\beta,x_i)}\left\{f(i-1,j-1,(\bar{\alpha},\bar{\beta}))\right\}
\eeq

Combining the two subcases in the case of $x_i=y_j$, we conclude that the recursive formula \eqn{eq31} is correct for the case $x_i=y_j$.

The proof is complete.
\hfill $\blacksquare$

\section{The Implementation of the Algorithm}
According to Theorem \ref{th1}, our algorithm for computing $f(i,j,(\alpha,\beta))$ is a standard 3-dimensional dynamic programming algorithm. By the recursive formula \eqn{eq31}, the dynamic programming algorithm for computing $f(i,j,(\alpha,\beta))$ can be implemented as the following Algorithm 1.

\begin{algorithm}
\caption{M-STR-IC-LCS}
{\bf Input:} Strings $X=x_1\cdots x_n$, $Y=y_1\cdots y_m$ of lengths $n$ and $m$, respectively, and a set of $d$ constraints $P=\{P_1,\cdots,P_d\}$ of total length $r$\\
{\bf Output:} The length of an LCS of $X$ and $Y$ including $P$
\begin{algorithmic}[1]
\STATE Build a keyword tree $T$ for $P$
\FORALL{$i,j$, $0\leq i\leq n, 0\leq j\leq m$}
\STATE $f(i,0,(0,0)) \leftarrow 0, f(0,j,(0,0)) \leftarrow 0$ \{boundary condition\}
\ENDFOR
\STATE $S \leftarrow \{(0,0)\}$ \{current set of states\}
\FOR{$i=1$ to $n$}
\FOR{$j=1$ to $m$}
\FOR{\textbf{each} $(\alpha,\beta)\in S$}
\IF {$x_i\neq y_j$}
\STATE $f(i,j,(\alpha,\beta)) \leftarrow \max\{f(i-1,j,(\alpha,\beta)),f(i,j-1,(\alpha,\beta))\}$
\ELSE
\STATE $\bar{\alpha} \leftarrow \delta(\alpha,x_i)$, $\bar{\beta}\leftarrow\gamma(\alpha,\beta,c)$, $s_{\bar{\beta}}\leftarrow s_{\beta}\bigcup O_{\bar{\alpha}}$
\STATE $f(i,j,(\bar{\alpha},\bar{\beta})) \leftarrow \max\{f(i-1,j-1,(\bar{\alpha},\bar{\beta})),1+f(i-1,j-1,(\alpha,\beta))\}$
\STATE $S \leftarrow S\bigcup\{(\bar{\alpha},\bar{\beta})\}$
\ENDIF
\ENDFOR
\ENDFOR
\ENDFOR
\RETURN $\max\limits_{0\leq i<t}\left\{f(n,m,(i,2^d-1)\right\}$
\end{algorithmic}
\end{algorithm}

In Algorithm 1, $T$ is the keyword tree for set $P$. The root of the keyword tree is numbered 0, and the other nodes are numbered $1,2,\cdots,t-1$ in their preorder numbering.
$\delta(\alpha,c)$ is the Aho-Corasick-next function defined in Definition \ref{df2}, which can be computed in $O(1)$ time. The function $\gamma(\alpha,\beta,c)$ is defined in Definition \ref{df3}, which can be computed in $O(d)$ time.
The variable $S$ is used to record the current states created. When the node of its output set is not empty is reached, a new state may be created. Therefore, in Algorithm 1, the current state set $S$ is extended gradually while the for loop processed. In the worst case, the set $S$ will have a size of $t2^d=O(2^dr)$, where $r$ is the total lengths of the constrained strings.
The body of the triple for loops can be computed in $O(d)$ time in the worst case. Therefor, the total time of
Algorithm 1 is $O(d2^dnmr)$. The space used by Algorithm 1 is $O(2^dnmr)$. In the case of the number of constraint strings is fixed, i.e. $d$ is a constant, our new algorithm for the M-STR-IC-LCS problem requires $O(nmr)$ time and space.

The number of constraints is an influent factor in the time and space complexities of our new algorithm. If a string $P_i$ in the constraint set $P$ is a proper substring of another string $P_j$ in $P$, then an LCS of $X$ and $Y$ including $P_j$ must also include $P_i$. For this reason, the constraint string $P_i$ can be removed from constraint set $P$ without changing the solution of the problem.
Without loss of generality, we can make the following two assumptions on the constraint set $P$.

\begin{asm}
There are no duplicated strings in the constraint set $P$.
\end{asm}

\begin{asm}
No string in the constraint set $P$ is a proper substring of any other string in $P$.
\end{asm}

If Assumption 1 is violated, then there must be some duplicated strings in the constraint set $P$. In this case, we can first sort the strings in the constraint set $P$, then duplicated strings can be removed from $P$ easily and then Assumption 1 on the constraint set $P$ is satisfied. It is clear that removed strings will not change the solution of the problem.

For Assumption 2, we first notice that a string $A$ in the constraint set $P$ is a proper substring of string $B$ in $P$, if and only if in the keyword tree $T$ of $P$, there is a directed path of failure links from a node $v$ on the path from the root to the leaf node corresponding to string $B$ to the leaf node corresponding to string $A$ \cite{1,9}.
For example, in Fig.1, there is a directed path of failure links from node 5 to node 7 and thus we know the string $ba$ corresponding to node 7 is a proper substring of string $aba$ corresponding to node 5.

With this fact, if Assumption 2 is violated, we can remove all proper substrings from the constraint set $P$ as follows.
We first build a keyword tree $T$ for the constraint set $P$, then mark all the leaf nodes pointed by a failure link in $T$ by using a depth first traversal of $T$. All the strings corresponding to the marked leaf node can then be removed from $P$. Assumption 2 is now satisfied on the new constraint set and the keyword tree $T$ for the new constraint set is then rebuilt. It is not difficult to do this preprocessing in $O(r)$ time. It is clear that the removed proper substrings will not change the solution of the problem.

If we want to compute the longest common subsequence of $X$ and $Y$ including $P$, but not just its length, we can also present a simple recursive backtracking algorithm for this purpose as the following Algorithm 2.

In the end of our new algorithm, we will find an index $\alpha$ such that $f(n,m,(\alpha,2^d-1))$ gives the length of an LCS of $X$ and $Y$ including $P$. Then, a function call $back(n,m,(\alpha,2^d-1))$ will produce the answer LCS accordingly.

\begin{algorithm}
\caption{$back(i,j,(\alpha,\beta))$}
{\bf Comments:} A recursive back tracing algorithm to construct the answer LCS
\begin{algorithmic}[1]
\IF{$i=0 \ \OR \ j=0$}
\RETURN
\ENDIF
\IF {$x_i=y_j$}
\IF {$f(i,j,(\alpha,\beta))=f(i-1,j-1,(\alpha,\beta))$}
\STATE $back(i-1,j-1,(\alpha,\beta))$
\ELSE
%\STATE $\tau \leftarrow 0$\\
\FOR{\textbf{each} $(\bar{\alpha},\bar{\beta})\in S$}
\IF {$\alpha=\delta(\bar{\alpha},x_i)$ \AND $\beta=\gamma(\bar{\alpha},\bar{\beta},x_i)$ \AND $f(i,j,(\alpha,\beta))=1+f(i-1,j-1,(\bar{\alpha},\bar{\beta}))$}
\STATE $back(i-1,j-1,(\bar{\alpha},\bar{\beta}))$
\PRINT $x_i$
\ENDIF
\ENDFOR
\ENDIF
\ELSIF{$f(i-1,j,(\alpha,\beta))>f(i,j-1,(\alpha,\beta))$}
\STATE $back(i-1,j,(\alpha,\beta))$
\ELSE
\STATE $back(i,j-1,(\alpha,\beta))$
\ENDIF
\end{algorithmic}
\end{algorithm}

Since the cost of $\delta(k,x_i)$ is $O(1)$ in the worst case, the time complexity of the algorithm $back(i,j,k)$ is $O(n+m)$.

Finally we summarize our results in the following Theorem.

\begin{theo}\label{th2}
For the two input sequences $X=x_1x_2\cdots x_n$ and $Y=y_1y_2\cdots y_m$ of lengths $n$ and $m$, respectively, and a set of $d$ constraints $P=\{P_1,\cdots,P_d\}$ of total length $r$, the Algorithms 1 and 2 solve the M-STR-IC-LCS problem correctly in $O(d2^dnmr)$ time and $O(2^dnmr)$ space, with preprocessing time $O(r|\Sigma|)$. In the case of the number of constraint strings is fixed, the Algorithms 1 and 2 for the M-STR-IC-LCS problem require $O(nmr)$ time and space.
\end{theo}

\section{Concluding Remarks}
We have suggested a new dynamic programming solution for the new generalized constrained longest common subsequence problem M-STR-IC-LCS. The new dynamic programming algorithm requires $O(d2^dnmr)$ time in the worst case. In the case of the number of constraint strings $d$ is fixed, our new algorithm for the M-STR-IC-LCS problem requires $O(nmr)$ time and space, and thus this is a polynomial time algorithm. If $d$ is not fixed, the time complexity $O(d2^dnmr)$ is still exponential in its expression. It is not clear whether there is an efficient algorithm in this case. We conjecture that our new algorithm is still polynomial even though $d$ is not fixed. We will investigate this issue further.


\begin{thebibliography}{23}

\bibitem{1}
Aho A.V., Corasick M.J., Efficient string matching: an aid to bibliographic search, \emph{Commun ACM} 18(6), 1975, pp. 333-340.

\bibitem{2}
Ann H.Y., Yang C.B., Tseng C.T., Hor C.Y., A fast and simple algorithm for computing the longest common subsequence of run-length encoded strings, \emph{Inform Process Lett} 108(11), 2008, pp.360-364.

\bibitem{3}
Ann H.Y., Yang C.B., Peng Y.H., Liaw B.C., Efficient algorithms for the block edit problems, \emph{Inf Comput} 208(3),2010, pp. 221-229.

\bibitem{4}
Apostolico A., Guerra C., The longest common subsequences problem revisited, \emph{Algorithmica} 2(1),1987, pp.315-336.

\bibitem{5}
Arslan A.N., Egecioglu O., Algorithms for the constrained longest common subsequence problems, \emph{Int J Found Comput Sci} 16(6), 2005, pp. 1099-1109.

\bibitem{6}
Blum C., Blesa M.J., L¨®pez-Ib¨¢nez M., Beam search for the longest common subsequence problem, \emph{Comput Oper Res} 36(12), 2009, pp. 3178-3186.

\bibitem{7}
Chen Y.C., Chao K.M., On the generalized constrained longest common subsequence problems, \emph{J Comb Optim} 21(3), 2011, pp. 383-392.

\bibitem{8}
Chin F.Y.L.,Santis A.D.,Ferrara A.L.,Ho N.L.,Kim S.K., A simple algorithm for the constrained sequence
problems, \emph{Inform Process Lett} 90(4), 2004, pp. 175-179.

\bibitem{9}
Crochemore M.,Hancart C., and Lecroq T., Algorithms on strings, Cambridge University Press, Cambridge, UK, 2007.

\bibitem{10}
Deorowicz S., Quadratic-time algorithm for a string constrained LCS problem, \emph{Inform Process Lett} 112(11), 2012, pp. 423-426.

\bibitem{11}
Deorowicz S., Obstoj J., Constrained longest common subsequence computing algorithms in practice, \emph{Comput Inform} 29(3), 2010, pp. 427-445.

\bibitem{12}
Farhana E., Ferdous J., Moosa T., Rahman M.S., Finite automata based algorithms for the generalized
constrained longest common subsequence problems, In: Proceedings of the 17th international conference
on string processing and information retrieval, SPIRE¡¯10, Los Cabos, Mexico, 2010, pp. 243-249.

\bibitem{13}
Gotthilf Z., Hermelin D., Lewenstein M., Constrained LCS: hardness and approximation. In: \emph{Proceedings of the 19th annual symposium on combinatorial pattern matching, CPM'08}, Pisa, Italy, 2008, pp. 255-262.

\bibitem{14}
Gotthilf Z., Hermelin D., Landau G.M., Lewenstein M., Restricted LCS. In: \emph{Proceedings of the 17th international conference on string processing and information retrieval, SPIRE'10}, Los Cabos, Mexico, 2010, pp. 250-257.

\bibitem{15}
Gusfield, D.,Algorithms on Strings, Trees, and Sequences: Computer Science and Computational Biology. Cambridge University Press, Cambridge, UK, 1997.

\bibitem{16}
Hirschberg D.S., Algorithms for the longest common subsequence problem, \emph{J ACM} 24(4), 1977, pp. 664-675.

\bibitem{17}
Hunt J.W., Szymanski T.G., A fast algorithm for computing longest common subsequences, \emph{Commun ACM} 20(5), 1977, pp. 350-353.

\bibitem{18}
Iliopoulos C.S., Rahman M.S., New efficient algorithms for the LCS and constrained LCS problems, \emph{Inform Process Lett} 106(1), 2008, pp. 13-18.

\bibitem{19}
Iliopoulos C.S., Rahman M.S., A new efficient algorithm for computing the longest common subsequence, \emph{Theor Comput Sci} 45(2), 2009, pp. 355-371.

\bibitem{20}
Iliopoulos C.S., Rahman M.S., Rytter W., Algorithms for two versions of LCS problem for indeterminate strings, \emph{J Comb Math Comb Comput} 71, 2009, pp. 155-172.

\bibitem{21}
Iliopoulos C.S., Rahman M.S., Vor¨¢cek M., Vagner L., Finite automata based algorithms on subsequences
and supersequences of degenerate strings, \emph{J Discret Algorithm} 8(2), 2010, pp. 117-130.

\bibitem{22}
Knuth D.E., Morris J.H.Jr, Pratt V., Fast pattern matching in strings, \emph{SIAM J Comput} 6(2), 1977, pp. 323-350.

\bibitem{23}
Maier D., The complexity of some problems on subsequences and supersequences, \emph{J ACM} 25, 1978, pp. 322-336.

\bibitem{24}
Peng Y.H., Yang C.B., Huang K.S., Tseng K.T., An algorithm and applications to sequence alignment with weighted constraints, \emph{Int J Found Comput Sci} 21(1),2010, pp. 51-59.

\bibitem{25}
Shyu S.J., Tsai C.Y., Finding the longest common subsequence for multiple biological sequences by ant colony optimization, \emph{Comput Oper Res} 36(1), 2009, pp. 73-91.

\bibitem{26}
Tang C.Y., Lu C.L., Constrained multiple sequence alignment tool development and its application to RNase family alignment, \emph{J Bioinform Comput Biol} 1, 2003, pp. 267-287.

\bibitem{27}
Tsai Y.T., The constrained longest common subsequence problem, \emph{Inform Process Lett} 88(4), 2003, pp. 173-176.

\bibitem{28}
Tseng C.T., Yang C.B., Ann H.Y., Efficient algorithms for the longest common subsequence problem with sequential substring constraints, \emph{J Complexity} 29, 2013, pp. 44-52.

\bibitem{29}
Wagner R., Fischer M., The string-to-string correction problem, \emph{J ACM} 21(1), 1974, pp. 168-173.

\bibitem{30}
Wang L., Wang  X., Wu Y., Zhu  D., A dynamic programming solution to a generalized LCS problem, \emph{Inform Process Lett} 113(1), 2013, pp. 723-728.

\end{thebibliography}
\end{document}